\documentclass{sig-alternate-05-2015}

\usepackage{graphicx}             
\usepackage{caption}
\usepackage{subcaption}
\usepackage{url}                  
\usepackage[latin1]{inputenc}     
\usepackage{nicefrac}             
\usepackage{indentfirst}          
\usepackage{amsfonts}
\usepackage{amssymb}
\usepackage{amsmath}
\usepackage[super, negative]{nth}
\usepackage{slashbox}
\usepackage[table]{xcolor}
\usepackage{multirow}
\usepackage{algorithm}
\usepackage{algpseudocode}
\usepackage{fancyvrb}
\usepackage{paralist}
\usepackage{xspace}

\newcommand{\framework}{REPETITA\xspace}
\newcommand{\repository}{{\tt https://github.com/svissicchio/Repetita}\xspace}

\newcommand{\myitem}[1]{\vspace*{0.07in}\noindent\textbf{#1}}
\newcommand{\afteritem}{\vspace*{0.07in}}

\newcommand{\remove}[1]{}

\newcommand{\ecmp}{\mathit{ecmp}}
\newcommand{\srpath}{\mathit{path}}

\def\sharedaffiliation{%
\end{tabular}
\begin{tabular}{c}}

\usepackage{etoolbox}
\makeatletter
\patchcmd{\maketitle}{\@copyrightspace}{}{}{}
\makeatother

\begin{document}

\title{REPETITA: Repeatable Experiments for Performance Evaluation of Traffic-Engineering Algorithms}

\numberofauthors{1}
\author{
\alignauthor
Steven Gay~\footnotemark[1] , Pierre Schaus~\footnotemark[1] , Stefano Vissicchio~\footnotemark[2] 
\sharedaffiliation
       \affaddr{
       	\footnotemark[1]\ \  Universite catholique de Louvain, 
       	\footnotemark[2]\ \  University College London
       	}\\
        \footnotemark[1]\  \tt{firstname.lastname@uclouvain.be},
        \footnotemark[2]\  \tt{s.vissicchio@cs.ucl.ac.uk}
}

\maketitle

\begin{abstract}
In this paper, we propose a pragmatic approach to improve reproducibility of
experimental analyses of traffic engineering (TE) algorithms, whose
implementation, evaluation and comparison are currently hard to replicate.
Our envisioned goal is to enable universally-checkable experiments of existing
and future TE algorithms.

We describe the design and implementation of \framework, a software framework
that implements common TE functions, automates experimental setup, and eases
comparisons (in terms of solution quality, execution time, etc.) of TE
algorithms.
In its current version, REPETITA includes
\begin{inparaenum}[(i)]
\item a dataset for repeatable experiments, consisting of more than 250 real
network topologies with complete bandwidth and delay information as well as
associated traffic matrices; and
\item the implementation of state-of-the-art algorithms for intra-domain TE with
IGP weight tweaking and Segment Routing optimization.
\end{inparaenum}
We showcase how our framework can successfully reproduce results described in
the literature, and ease new analyses of qualitatively-diverse TE algorithms.
We publicly release our REPETITA implementation, hoping that the community will
consider it as a demonstration of feasibility, an incentive and an initial code
basis for improving experiment reproducibility: Its plugin-oriented architecture
indeed makes \framework easy to extend with new datasets, algorithms,
TE primitives and analyses.
We therefore invite the research community to use and contribute to our released
code and dataset.
\end{abstract}

\section{Introduction}
Reproducibility is a cornerstone of the scientific method, and a strongly
desirable principle for any exact-science discipline -- computer science
included.
By enabling third parties to check experimental results, it indeed ensures the
validity of published contributions and stimulates new ones, ultimately
providing a better, community-shared understanding of the state of the art.

Unfortunately, reproducibility of research in networking
exhibits huge room for improvement.
Despite recent enablers of reproducibility like network
simulators~\cite{Weingartner:2009}, virtualization tools~\cite{mininet-conext12}
and testbeds~\cite{planetlab-03}, articles still often come with a rough
description of the performed experiments, and limited or no support (code,
dataset, etc.) to reproduce the presented evaluation.
As a consequence, reviewers have to believe the claimed results.
Also, researchers willing to compare new proposals with the state of the art
must re-implement technical contributions (to the best of their understanding)
and re-create experiments from scratch:
As confirmed by the few initiatives on reproducing research results like the
Stanford CS244 course~\cite{stanford-cs244-ccr},
this is a challenging, time-consuming task, that further provides no guarantees
of achieving a fair comparison against the original proposal.

We believe that the networking community should find more effective ways to
ease support for reproducibility, at least for a significant part
of the experiments in every publication.

Of course, achieving reproducibility is not straightforward.
In this work, we focus on traffic engineering (TE).
For TE, a primary challenge consists in having
access to both experimental settings, evaluation datasets, and
evaluated algorithms, which unfortunately is costly if not
impossible.
The original experimental setting is generally hard to re-create: for example,
testing infrastructures (e.g., servers, network
equipment and software) are often not shared among researchers, may not be affordable
for all research institutes, and tend to quickly become unsupported over time
(e.g., because of new software releases).
Also, evaluations on real, production-network data involve privacy
concerns, as they contain sensitive information that 
can rarely be disclosed.
Finally, releasing code (e.g., for algorithms, system prototypes, etc.) implies
a significant effort that often comes with little benefits for authors.

We propose an approach based on \textit{repeatability},
a looser form of reproducibility focused on reproducing a given set of
experiments in a local setting
(e.g., on a different server with respect to the one used in the
original papers).
By itself, repeatability avoids the main roadblocks deriving from rebuilding the
original experimental setting.

To tackle the other challenges, we argue for more effective practices and tools
that facilitate repeatability.
As a concrete step in this direction, we publicly release\footnote{see
\repository} \framework, a software platform that we have also used to develop
our own research contributions~\cite{srls-infocom17}.
\framework eases repeatable evaluation of TE algorithms: It automates most of the
technical operations involved (topology pre-processing, evaluation setup,
analyses, etc.) and enables users to run experiments on more than 250 topologies
(with 5 synthetic traffic matrices each) with a single command-line instruction.

Admittedly, \framework is not a silver bullet for any possible evaluation on TE.
Its dataset can be improved by adding more (e.g., different types and more
updated) real topologies of networks, as well as real traffic data.
The implemented algorithms mainly focus on optimization with two basic
primitives (IGP and Segment Routing) -- even if the configuration of explicit
paths, like MPLS tunnels and OpenFlow-installed paths, is also supported.
The evaluation metrics are far from covering the full spectrum of possible
evaluations: they mainly focus on bandwidth optimization (e.g., in contrast with
recent TE systems like~\cite{B4:2013,swan-sigcomm13}). Even the type of possible
experiments is limited, as it restricts to the static optimization of forwarding
paths, with no knowledge of transient states and system dynamics.

Nevertheless, we believe that no repeatability-oriented approach can be
successful if it does not involve a community effort and commitment.
We designed and implemented \framework to be extendable, so that it can benefit
from the community push that we argue for.
For example, its plugin-oriented architecture enable to easily enrich the amount
of supported algorithms and primitives (e.g., underlying protocols and router
capabilities).

Our framework is also easy to use.
We hope that usability can (i) provide researchers with immediate incentives and
technical means to balance the practical value of experiments on private datasets
with the scientific importance of reproducible ones, in future publications on
TE; (ii) spur the creation of benchmarks for TE algorithms, ideally merging
contributions from the wide research and operator communities; (iii) reviewers
and external readers to check evaluation results; and (iv) stimulate approaches
that support repeatable evaluations of research contributions in other areas.

In the following, we describe how \framework fills a gap in the state of the art
(\S\ref{sec:related-work}), how it is designed (\S\ref{sec:framework}) and
implemented (\S\ref{sec:implementation}), and how it can be used for both
repeating experiments and facilitating new insightful analyses
(\S\ref{sec:eval}).

\section{What do we Lack?}\label{sec:related-work}

\myitem{Many networking contributions are hard to reproduce.}
Position papers~\cite{reproduciblecs-peng11},
methodologies~\cite{recomp-manifesto} and repositories of
data~\cite{beasley1990or,xcsp3} have been published to facilitate experiment
reproducibility in several areas of computer science.

In our experience, however, most networking contributions are hard to reproduce
and fairly compare against.
For instance, the majority of papers on wide-area traffic engineering (from
earlier on IGP weight optimization~\cite{Fortz_Internet:2000} to more
recent approaches based on MPLS~\cite{elwalid2001mate,Kandula:2005-texcp},
SDN~\cite{B4:2013,swan-sigcomm13} and SR~\cite{belllabs-infocom15})
present experiments made with proprietary code, often on a restricted dataset of
private networks.

\myitem{The networking community is pushing towards open-source code.}
Software Defined Networking (SDN) originated from the definition of
OpenFlow~\cite{mckeown2008openflow}, an open-source interface to network
devices.
Consistently, many SDN proposals, from projects about SDN programming languages
(e.g.,~\cite{frenetic-11}) to those on network controllers
(e.g.,~\cite{nox-08}), have publicly released the produced code -- an
encouraging step towards reproducibility. A few recent works also consider the
problem of fairly comparing open-source proposals.
Prominently, they focus on tools to benchmark SDN controllers
(e.g.,~\cite{sdload-14,hvbench-16}) and OpenFlow switches
(e.g.,~\cite{oflops-12}), under specific network conditions or workloads.
While close in spirit with those contributions, this paper has a different scope
(traffic engineering), that also
comes with its own challenges like working around confidentiality of data
(network topologies, traffic matrices, etc.) used in experiments.

\myitem{We still miss approaches and tools that make reproducibility easy.}
The literature describes a pletora of tools that potentially enable repeatability,
including simulators (e.g.,~\cite{Weingartner:2009}), emulation platforms
(e.g.,~\cite{mininet-conext12}) and testbeds (e.g.,~\cite{planetlab-03}).
They provide the low-level means to perform network experiments at scale, but
everything else (network setup, approaches to be evaluated, experimental
configuration, etc.) must be added on top of them.
As such, those tools make repeatability just possible.

We however believe that the networking community still lacks proven methodologies, tools and practices that make reproducibility not only possible, but also easy and affordable for researchers (and reviewers).

For example, a past proposal to enable TE experiments in non-SDN
networks is TOTEM~\cite{totem-comcom}.
TOTEM defines a set of TE algorithms, but no standard experiment or analysis on
them.
Further, the code is not maintained since years.
Hence, newer algorithms and support for cutting-edge technologies (from
OpenFlow to Segment Routing) are not implemented.
Even worse, it is unclear how to support such algorithms and technologies in
TOTEM, since the tool is mainly a collection of heterogeneous scripts only
sharing input and output format.
By releasing an extendable platform that includes recent TE approaches and
automates all the main operations to run repeatable experiments,
we hope to overcome the difficulties of tools like TOTEM and provide
immediate incentives for reproducibility in the TE field.

\section{Repeatability Made Simpler}\label{sec:framework}
In this section, we overview \framework.
We describe its design (\S\ref{subsec:framework-arch}), workflow
(\S\ref{subsec:framework-basics}), and key benefits
(\S\ref{subsec:framework-benefits}).

\subsection{Designing \framework}\label{subsec:framework-arch}

\begin{figure}[]
	\centering
	\includegraphics[width=\columnwidth]{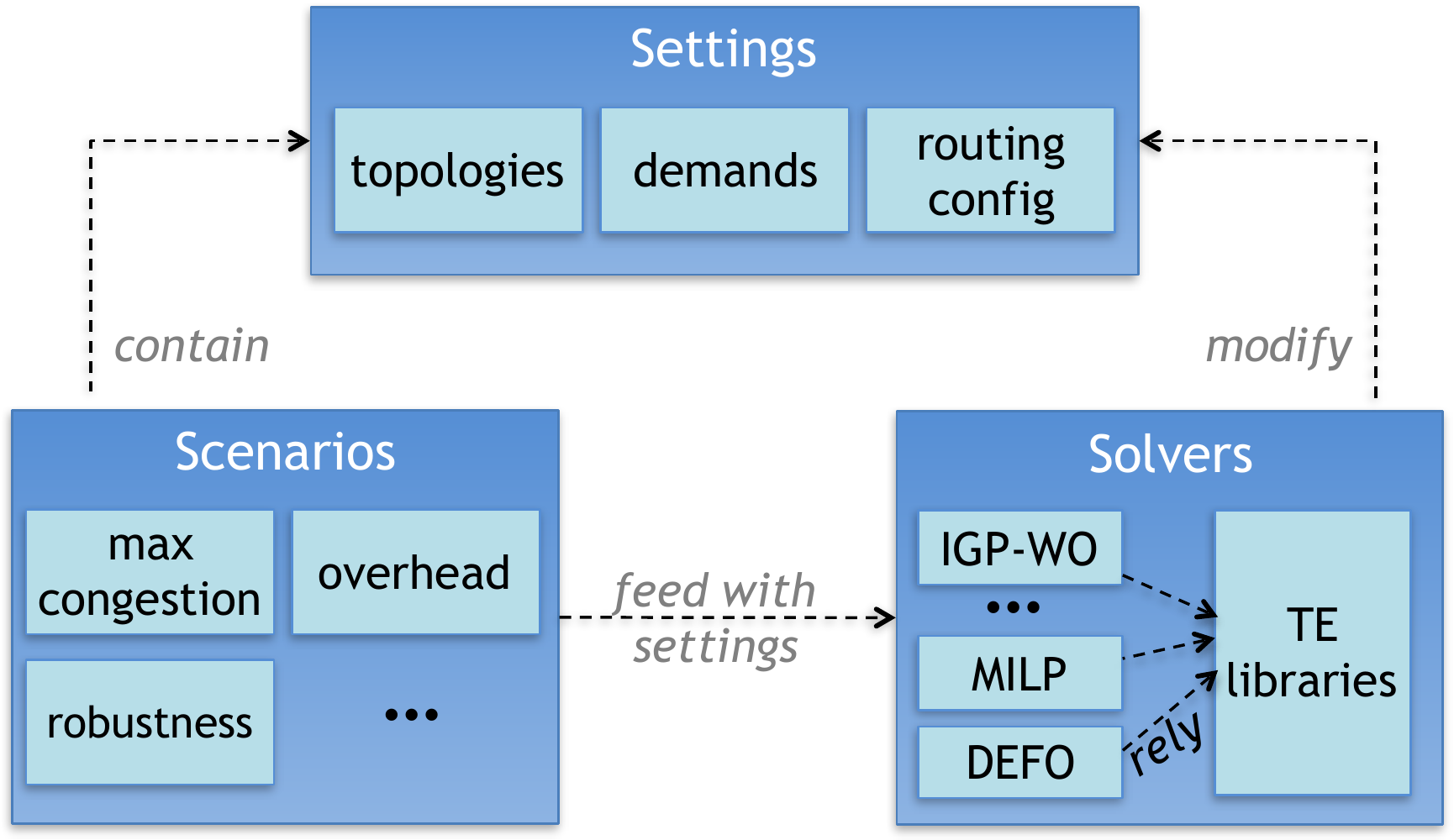}
	\caption{\framework architecture.}
	\label{fig:architecture}
\end{figure}

As shown in Figure~\ref{fig:architecture}, \framework design is based on
modeling and separating three main concepts: settings, solvers, and scenarios.
Altogether, they define the \textit{experiments and analyses} that are performed
in a run of \framework.

\myitem{Settings.} A setting represents the problem instance considered in an
experiment. It therefore groups all the input data needed to run an
experiment.
In the TE case, settings include topologies, traffic matrices, and routing
configurations.
A \textit{topology} represents a network and contains all its characteristics
that are relevant for traffic engineering. We identify nodes, links, link
capacities and delays as a basic set of such characteristics.
A \textit{traffic matrix} encompasses data (sources, destinations and volumes)
about traffic flows assumed to traverse the network in a given experiment.
A \textit{routing configuration} contains additional information to compute
forwarding paths on the given topology and traffic matrix. Primarily, it
includes link weights for shortest-path computations done by current
intra-domain routing protocols (called IGPs).
Also, it encompasses other protocol-specific configurations, like Segment
Routing settings as well as explicit paths reflecting configured MPLS tunnels or
OpenFlow rules.

\myitem{Solvers.} A solver is an object under test. In our case, a solver is a
TE algorithm that given a setting, updates the routing configuration inside the
setting for the corresponding topology and traffic matrix.
This computation typically aims at optimizing network
performance, e.g., maximizing bandwidth, avoiding congestion or reducing
delay.
Solvers are pieces of code that are either standalone or integrated in
\framework. In the latter case, \framework offers \textit{TE libraries} 
implementing building blocks like computation of IGP shortest paths, and the
calculation of link utilizations induced by a given routing configuration.

\myitem{Scenarios.} A scenario models the dimensions along which the output of
different solvers is analyzed.
In other words, scenarios define analyses of interest for \framework users, and
provide automated support to perform such analyses\footnote{\small Note that
	scenario has a different meaning in TOTEM where it only defines events
	occurring during an experiment.}.
For example, a scenario can define how good a TE algorithm (solver) is at
optimizing bandwidth on a given network and traffic matrix (setting).
In this sense, scenarios implement the abstractions (i.e., the evaluation
metrics) that ensure comparability between different solvers.

\afteritem

We believe that \framework architecture can fit a broader class of networking
problems, as for example congestion control (where solvers are congestion
control schemes, settings encompass traffic rates and network conditions, and
scenarios evaluate performance metrics as throughput) and security (where
solvers are defense techniques, settings models networks, targets and attacks,
and scenarios quantify success rate of solvers as well as consumed resources and
time).

\subsection{Running \framework}\label{subsec:framework-basics}
\framework provides the software infrastructure that links together the building
blocks discussed in \S\ref{subsec:framework-arch}.

We now list the steps sequentially performed by \framework to carry out a
(repeatable) experiment.
Note that the framework also provides direct access to the functions used in
those steps, hence it can be used as a software library (e.g., to solve
multi-commodity flow problems~\cite{shahrokhi1990maximum} and calculate lower
bounds for any TE algorithm).

The first step consists in processing the input.
\framework parses topology and demands files using custom parsers: Those parsers
create setting objects from files of tabular format, where every line describes
an attribute (e.g., IGP weight, traffic volume, etc.) of a network element
(node, link or demand).
Once such files are parsed, \framework instantiates solvers and scenarios
required in the experiments to be performed.

Scenarios are then asked to run.
For the simplest possible scenario, this method call translates into passing the
configured setting to the solver stored in the scenario, and reporting the
differences (e.g., in terms of maximum link utilization) between the routing
configuration before and after the solver's optimization.
More advanced scenarios can also modify the original settings or create new ones.
For example, consider a scenario that requires to evaluate what happens in the
case of any single link failure (like the robustness analysis described in
\S\ref{subsec:impl-scenarios}).
Such a scenario instantiates many new settings, each corresponding to a single
link failure on the initial topology -- and runs the configured solver on all the
new settings.

Scenarios also implement two key functions.
First, they enforce that every solver's execution terminates after a
(potentially infinite) time bound, e.g., stopping it after the given time bound
and extracting the best routing configuration computed by the solver so far.
Second, scenarios process the routing configurations returned by the solvers.
Namely, they translate each of those configurations into paths, map the input
traffic matrix to the computed paths, and evaluate metrics like link utilization
or routing overhead.
Such a translation leverages functions implemented in the \framework TE
libraries, for independence from solver implementations.

Results of scenarios' post-processing are finally reported on the screen (by
default) or on a pre-configured output file.

\subsection{Benefiting from \framework Design}\label{subsec:framework-benefits}

\framework is designed to achieve usability and extensibility.
\noindent Technical documentation, instructions to run
experiments and examples to add new solvers are provided at \repository.

\myitem{\framework is easy to use.}
Users can run the implemented solvers on supported settings and scenarios by
executing the framework from the command line.
The following working example instructs \framework to run a solver called
\texttt{defoCP}, once, for 1 second, on the Abilene network.

\begingroup
\small
\begin{center}
	\begin{Verbatim}[frame=single]
repetita -graph Abilene.graph -demands Abilene.demands
         -solver defoCP -t 1 -scenario SingleSolverRun
	\end{Verbatim}
\end{center}
\endgroup

\noindent A few more CLI options enable users to adjust parameters like the
output file name (\framework results are printed to the CLI standard output by
default).

\myitem{\framework is easy to run on already implemented solvers.}
For ease of adoption, we do not require solvers to be implemented within our
framework.
In contrast, our framework can run external solvers (written in any programming
language) as long as they are packaged as executables that take as input a
network and a traffic matrix. To add such solvers, it is sufficient to
specify information about them in a textual file (i.e.,
external\_solvers/solvers-specs.txt) inside the \framework directory.
Figure~\ref{fig:externalconf_example} reports a snippet of such file, displaying
the description of an example solver (getRandomPaths.py) written in
Python.
As shown in the figure, limited information must be specified: general data like
the name (to refer the solver within \framework) and the optimization objective;
specification of the CLI commands to run the executable and extract the time
taken by its last run; a description on how to interpret the output of the
executable. More details on the file format are provided in the file itself.

\begin{figure}
	\begingroup
	\small
	\begin{center}
		\begin{Verbatim}[frame=single]
// general information about the solver
name = randomTunnels
optimization objective = 'undefined'

// how to run the external solver
run command = python external_solvers/getRandomPaths.py
$TOPOFILE $DEMANDFILE $OUTFILE

// how to interpret the output of the last solver's run
optimization effect = setExplicitPaths
field separator = '; '
key field = 0
value field = 2

// how to get the time taken by the last solver's run
gettime command = cat $OUTFILE | grep 'execution time'
| awk -v FS=': ' '{print $2}'
		\end{Verbatim}
	\end{center}
	\endgroup
	\vspace{-0.4 cm}
	\caption{Configuration of an external solver (in Python), added to
		\framework.}
	\label{fig:externalconf_example}
	\vspace{-0.3cm}
\end{figure}

\myitem{\framework is easy to extend.}
Its architecture supports separate evolution (e.g., modification or replacement)
of the dataset, the solvers and the evaluation analyses.
Topologies and traffic data are controlled with a command-line parameter, that
specifies the directory containing them.
Additional analyses can be implemented by defining new scenario objects in the
\framework code.
Beyond using external solvers, new algorithms can also be integrated within the
framework.
Two options are viable, as detailed on the \framework Web page.
The first option is to re-implement the original solver using the \framework TE
libraries, as we did for our 156-line long implementation of the IGP weight
optimization heuristic based on~\cite{Fortz_Internet:2000}.
A more lightweight alternative consists in adding the new solvers, unmodified,
along with solver-specific wrappers, i.e., software objects that convert input
and output of method calls between the framework and the original solver.
We used the latter option to support the original DEFO solver~\cite{defo-sig15}
(in Scala) in our current \framework implementation.

\section{Repeat, for Real!}\label{sec:implementation}

We now describe settings (\S\ref{subsec:impl-settings}), solvers
(\S\ref{subsec:impl-solvers}) and scenarios (\S\ref{subsec:impl-scenarios})
supported by our Java-based implementation.

\subsection{Topologies and Traffic Matrices}\label{subsec:impl-settings}

We first detail how settings are built, and why.

\myitem{We collected realistic topologies}
used in the evaluation of previous TE papers.
We relied on 3 public data sources: the Rocketfuel
project~\cite{rocketfuel-ton04}, the synthetic topologies used in
DEFO~\cite{defo-sig15}, and the Internet Topology Zoo \cite{knight2011internet}.
The former two sources provide a limited number of topologies, either
inferred from Internet measurements (Rocketfuel) or artificially generated to
stress-test algorithm scalability (DEFO).
Internet Topology Zoo gathers real topologies (often WANs)
as reported by official Web sites of commercial companies.

\myitem{We post-processed the available topologies to have a complete dataset.}
Original topologies in the Internet Topology Zoo dataset were incomplete
in several cases.
We completed them as follows.
First, we kept only the largest connected component of any disconnected
topology.
Second, for topologies where no link capacity is specified, we set the same
value on all links.
Third, when capacity information are specified only for a subset of
network links, we set the capacity of any link with missing capacity as equal to
the average capacity across all links.
Further, we normalized link capacities to avoid too large variations (e.g., Kbps
access links versus Gbps core links) leading to unavoidable traffic bottlenecks
on which all TE algorithms would perform equally bad.
In particular, we impose that the value of any link capacity is not less than
$1/20$-th of the largest link capacity in the topology.

\myitem{We considered $3$ heuristics to assign IGP weights.}
Intra-domain routing protocols (i.e., IGPs) are based on shortest-path
computation.
We therefore need IGP weights of all the links in every topology to compute pre-
and post-TE paths.
Unfortunately, the Internet Topology Zoo dataset does not contain this information.
Our framework implementation then uses
$3$ standard heuristic assignments of IGP weights.
The first heuristic assigns unary weight to all links: It is motivated by
simplicity, and by the intuition that unitary weights may provide the most flexibility for protocols 
built on top of IGP (like Segment Routing~\cite{defo-sig15}).
The second heuristic sets IGP weights as inversely proportional to link
capacities, i.e., reflecting a default that has been classically adopted by
vendors\footnote{\small e.g., see
	http://www.cisco.com/c/en/us/support/docs/ip/open-shortest-path-first-ospf/7039-1.html\#t6}.
The third option is to consider weights optimized for a given traffic matrix, i.e.,
according to our IGP-WO solver (see \S\ref{subsec:impl-solvers}).

\myitem{We synthesized traffic matrices.}
We adopted a randomized gravity model that has been shown to generate realistic
traffic matrices~\cite{syntheticTMs-ccr05} and is often used to evaluate TE
algorithms (\cite{Kandula:2005-texcp},\cite{defo-sig15},...).
We generated $5$ traffic matrices per topology to have some diversity while
keeping the dataset manageable\footnote{\small For a couple of topologies, we
	could have also added old traffic matrices (e.g.,~\cite{geantTM-ccr06}): We
	decided not to do so, in order to have the same number and type of matrices for
	all topologies.}.
We scaled the traffic matrices so that the maximally utilized link is loaded at
90\% of its capacity in the optimal solution of the corresponding
multi-commodity flow problem (using the \framework LP described in
\S\ref{subsec:impl-solvers})
The maximal link utilization is thus larger or equal than 90\% in any real TE
solution.
We are well aware that operators (especially in WANs) do not typically run
production networks at a so high utilization rate. However, we applied this
scaling factor to build cases where very effective bandwidth-optimizing TE
algorithms are absolutely needed -- cases for which recent TE approaches in
private networks even argue~\cite{B4:2013,swan-sigcomm13}.
To match more common settings,
traffic-matrix values can be divided by a constant factor (e.g., $3$ for an
optimal maximum link utilization of $30\%$).

\subsection{TE Algorithms}\label{subsec:impl-solvers}

\framework includes a linear program that computes a \textbf{baseline} for
bandwidth-optimization algorithms.

\myitem{Linear Program (LP), solving the multi-commodity flow problem.}
The LP provides a theoretical lower bound for the classic TE goal of
minimizing the maximum link utilization.
It is indeed allowed to arbitrarily split traffic among any possible path,
and ignores IGP weights.
We used the LP for scaling the traffic matrices in our datasets (as just
described), and as a baseline in our experiments (see \S\ref{sec:eval}).
We report the LP formal definition in the Appendix.

\afteritem

In addition, \framework currently includes three \textbf{solvers}.
To show the possibility to implement algorithms based on
qualitatively-different primitives, we encoded a weight optimization algorithm
as well as recent proposals based on the newer Segment Routing (SR) protocol. To
ensure practicality of the implemented algorithms and apples-to-apples
comparison, all the solvers assume that traffic is equally split among all the
paths used from any traffic source to any destination -- a feature called
even load balancing.

\myitem{IGP Weight Optimization (IGP-WO), based on~\cite{Fortz_Internet:2000}.}
Given a traffic matrix and a topology, this algorithm runs a local search to
find IGP weights that minimize the load on the maximally-utilized link.
We implemented an approach based on~\cite{Fortz_Internet:2000} in 156 lines of
code by leveraging \framework TE libraries (e.g., for shortest-path
computation).

\myitem{Mixed Integer Linear Programming (MILP) optimization, inspired
	by~\cite{belllabs-infocom15}.}
We use standard optimization tools (i.e., the Gurobi
optimizer\footnote{\small see www.gurobi.com}) to solve mixed integer linear programs
inspired by the first work on SR-based TE, from Bhatia et
al.~\cite{belllabs-infocom15}.
As in~\cite{belllabs-infocom15}, our formulation admits that 2 IGP shortest
paths are stitched together, sharing a common node or \textit{detour}, thanks to
SR.
In contrast to~\cite{belllabs-infocom15}, though,
we removed the assumption that traffic flows could be split with arbitrary
ratios at the network ingress -- as none of the other algorithms relies on the
same assumption.
We also reduced the number of variables, using a modified node-link
formulation~\cite{pioro2004routing}, lowering the consumed memory.
We detail our MILP model in the Appendix.

\myitem{Constraint Programming heuristics (DEFO), as proposed in~\cite{defo-sig15}.}
DEFO~\cite{defo-sig15} is a Constraint Programming heuristic designed to trade
optimality for time efficiency and scalability.
This algorithm is heuristic in two ways.
First, it does not completely explore the solution space but iteratively samples
it using a randomization technique called Large Neighborhood Search.
Second, at every iteration, it greedily selects demands (from the biggest to the
smallest) and an unconstrained number of detours per demand.
We plugged DEFO in \framework by implementing a software wrapper for the
original DEFO code, as described in \S\ref{subsec:framework-benefits}.

\subsection{Scenarios}\label{subsec:impl-scenarios}
\framework readily supports the following scenarios.

\myitem{Max-congestion Analysis.}
This scenario computes the maximum link utilization of paths returned by
solvers. It therefore evaluates the quality of solutions found by input solvers
with respect to a classic traffic-engineering goal (see,
e.g.,~\cite{Fortz_Internet:2000}), i.e., minimization of maximal link
utilization.

\myitem{Overhead Analysis.}
Primitives like segment routing enables more flexibility in the choice of
forwarding paths at the cost of additional overhead (e.g., information added to
packets and router configuration to implement detours).
This scenario quantifies such overhead by keeping track of
changed IGP weights, traffic demands re-routed with SR, and modified explicit
paths.

\myitem{Robustness Analysis.}
\framework finally allows users to evaluate the robustness of a TE configuration
with respect to failures.
For any given topology and routing configuration (returned by a solver), the
robustness analysis scenario evaluates how the maximum link utilization
changes after the failure of every link that does not disconnect
the network. To this end, it recomputes the paths associated to the given
routing configuration (e.g., IGP weights or SR paths) on all the topologies
resulting from removing any single link from the original network.

\begin{figure*}[]
	\centering
	\begin{subfigure}[t]{.65\columnwidth}
		\includegraphics[width=\textwidth]{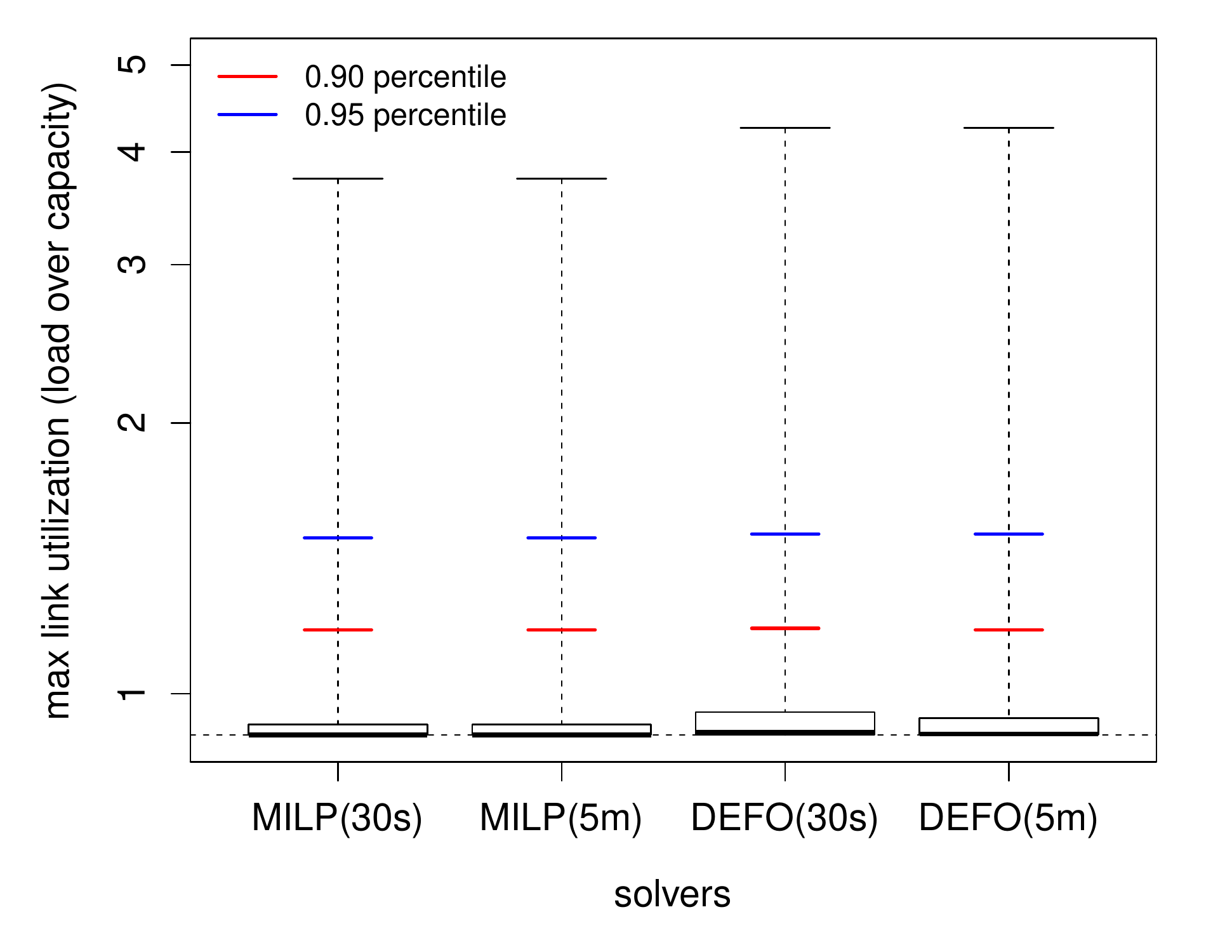}
		\caption{Small topologies (<30 nodes).}
		\label{fig:load-small}	
	\end{subfigure}
	\begin{subfigure}[t]{.65\columnwidth}
		\includegraphics[width=\textwidth]{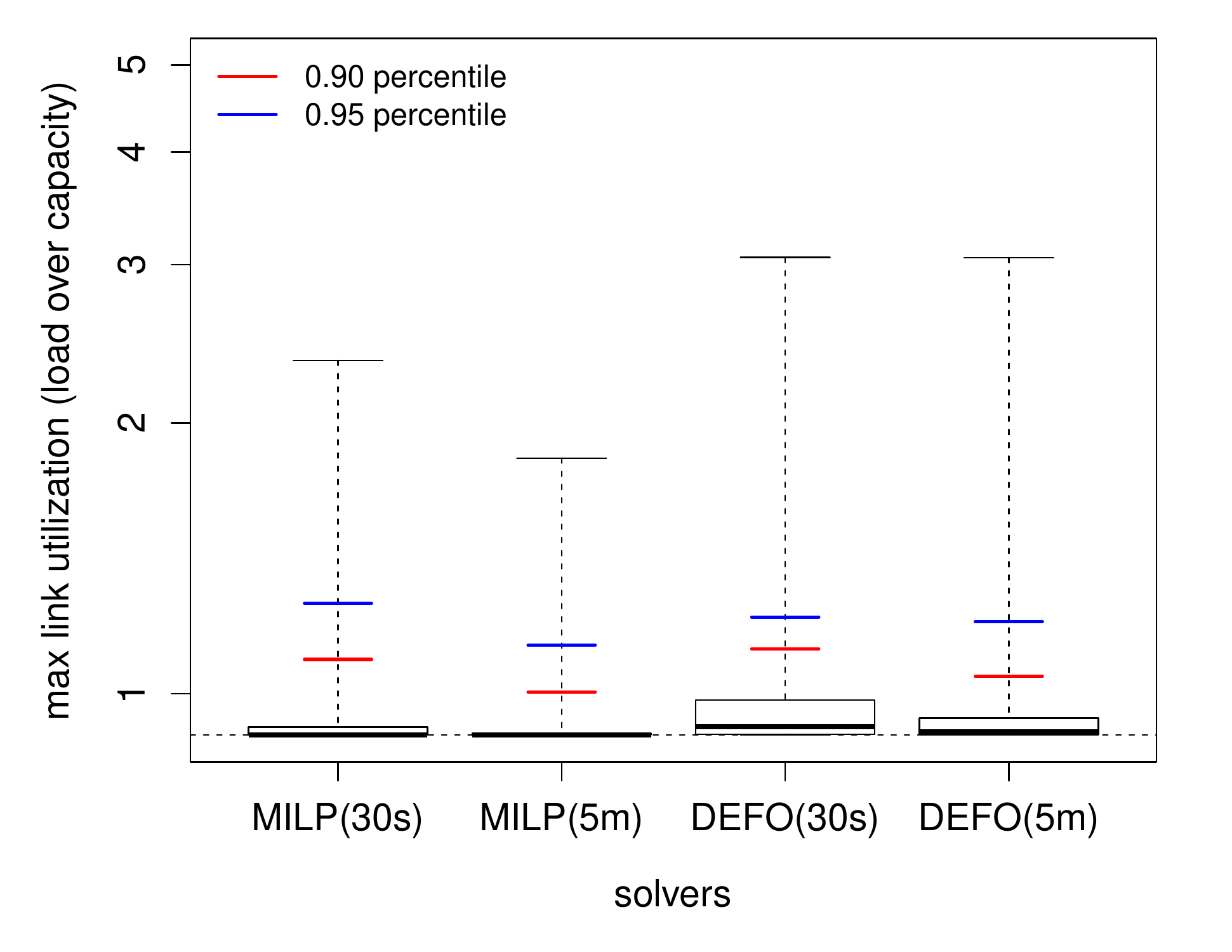}
		\caption{Medium topologies.}
		\label{fig:load-medium}
	\end{subfigure}
	\begin{subfigure}[t]{.65\columnwidth}
		\includegraphics[width=\textwidth]{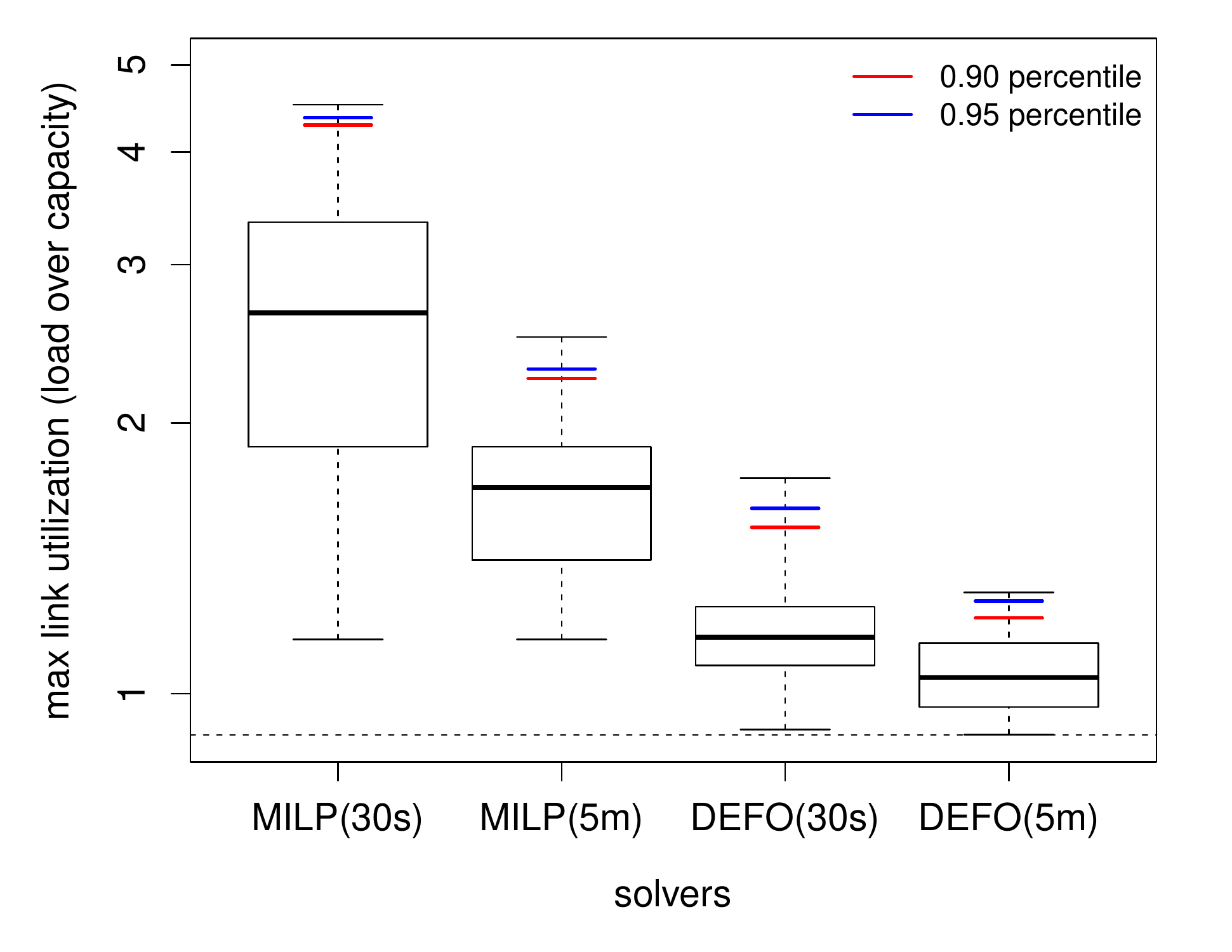}
		\caption{Large topologies (>100 nodes).}
		\label{fig:load-large}
	\end{subfigure}
	\caption{Max-congestion analysis results obtained with \framework. Top and bottom of the boxes represents $25$-th and $75$-th percentiles, the thick line in the boxes the median, and top and bottom whiskers minimum and maximum values, respectively. The dashed horizontal line is the theoretical lower bound, computed by \framework.}
	\label{fig:boxplot-load}
	\vspace{-1em}
\end{figure*}

\section{Repetita Iuvant\protect\footnote{\small L\MakeLowercase{atin for ``repeating does good''}}}
\label{sec:eval}
We ran our \framework implementation
using a 40-core Intel(R) Xeon(R) 3.10GHz machine with 128GB
memory. We limited multithreading of any given solver to 4 cores.
The used JVM is OpenJDK version 1.8.0\_91.
Experimental results are summarized in the following.

\myitem{\framework eases reproduction of published evaluations.}
We reproduced experiments described in~\cite{defo-sig15},
	running each of them $20$ times.
Our results confirmed what reported in the original
publication, with minimal differences (at most $3\%$) in the maximal link utilization
that should be ascribed to the randomized nature of the DEFO algorithm.

\myitem{\framework enables new, large-scale comparisons, not provided in the original
	papers.}
We compared MILP and DEFO performance on all the Internet Topology Zoo
networks with link weights assigned as inverse of their capacity.
We ran both solvers with two time limits (30 and 300 seconds), for a total of
5,200 experiments (1,300 settings per solver with a given time limit).
This is an original head-to-head comparison between the two algorithms.

Those experiments are meant to be a \textit{demonstration} of how our framework
can be used to evaluate different TE algorithms according to several metrics.
The evaluated solvers are indeed based on different principles: MILP looks for
optimal solutions, DEFO trades optimality guarantees for speed and scalability.
The following results quantify the tradeoffs achieved by those two algorithms.

First, we evaluated the quality of solutions returned by MILP and DEFO, as shown
in Figure~\ref{fig:boxplot-load}.
In few cases (about 10\%), both the solvers are quite far from the theoretical
lower bound, likely because of intrinsic limitations of segment routing (e.g.,
SR cannot force traffic on any possible path) and the assumed flow
unsplittability (for practicality).
The remaining experiments highlight fundamental differences between the
considered algorithms, mostly depending on the size of input topologies.
For small and medium topologies (Figures~\ref{fig:load-small}
and~\ref{fig:load-medium}), both solvers are within $10\%$ of the
theoretical lower bound, most of the time.
Nevertheless, MILP is closer to the theoretical optimum than DEFO, as expected
since the latter implements a heuristic approach.
For bigger topologies (Figure~\ref{fig:load-large}), however, DEFO definitely
outperforms MILP, as the latter often needs much more time to optimize SR
paths.
Also, the larger variation of results suggests
that the allowed execution time is more critical for MILP than for
DEFO, as the latter tends to converge faster to good TE solutions.

We also ran the overhead analysis for the two solvers.
Table~\ref{tab:overhead} stresses the huge difference between the high overhead
needed by MILP (to reach optimal traffic spread), and the little one induced by
DEFO (which moves far less flows).

\begin{table}[h]
	\centering
	\footnotesize
	\begin{tabular}{@{}lclclclcl@{}}
		\hline
		\emph{Experiments} && \emph{25\%} && \emph{Median} && \emph{75\%} && \emph{95\%} \\
		\hline
		MILP (30s) && 42.7\% && 56.6\% && 65.9\% && 75.2\% \\
		MILP (5m) && 42.7\% && 56.9\% && 66.1\% && 75.2\% \\
		DEFO (30s) && 1.1\% && 2.7\% && 6.1\% && 16.2\% \\
		DEFO (5m) && 1.6\% && 4.2\% && 8.8\% && 20\% \\
		\hline
	\end{tabular}
	\caption{Overhead comparison performed with \framework. The table reports the
		percentage of demands re-routed by MILP and DEFO, in the
		same experiments as in Figure~\ref{fig:boxplot-load}.}
	\label{tab:overhead}
	\vspace{-14pt}
\end{table}

\myitem{\framework facilitates new analyses.}
We extended the comparison between MILP and DEFO with metrics not
considered in previous papers.
We ran the robustness analysis scenario to assert whether solutions
computed with a congestion minimization focus are robust to failures.
We provided MILP and DEFO with a 300s time limit, and collected their
output for all topologies and one demand file per topology.
The robustness analysis scenario then simulated single link failures and
compared the maximum link utilization for each solution with their respective
theoretical lower bound.
On a total of $17,144$ possible link failures (across all networks), congestion 
appeared 
in $9,489$ cases for DEFO paths, and in $11,174$ MILP cases.
In contrast, \framework lower bound algorithm could not avoid congestion
in only $3,289$ cases.
This observation opens an interesting question on whether and to what
extent SR can be used to optimize both link utilization during normal operation
while simultaneously providing some guarantees in the case of failures.

\myitem{\framework motivates new research questions}, by showing the
impact of TE algorithms in specific settings. In particular, our experiments
highlight two aspects.

First, they confirm the potential relevance of TE algorithms whenever links are
run at high utilization. For example, Figure~\ref{fig:load-large} shows huge
differences in the ability to quickly remove congestion (as for online TE)
between two algorithms, despite relying on the same primitive (SR). This
observation complements the more system-oriented perspective taken by recent
works~\cite{B4:2013,swan-sigcomm13}, focused on how to robustly extract input
data and try to enforce TE decisions.

\begin{figure}[t]
	\vspace{1em}
	\centering
	\includegraphics[width=0.9\columnwidth]{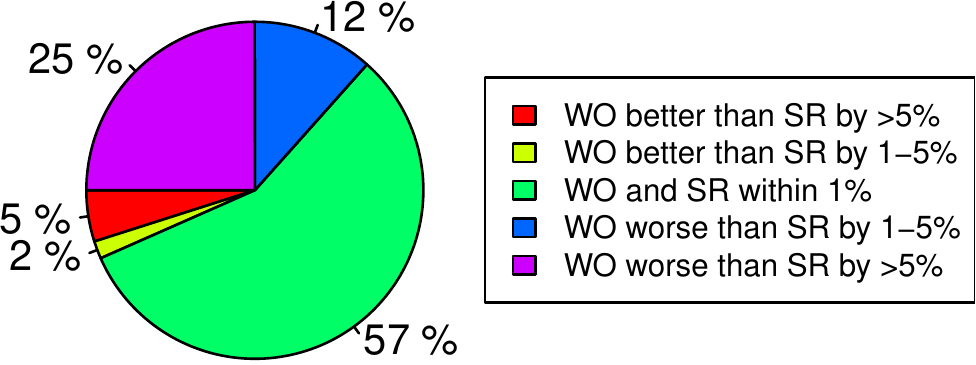}
	\caption{Systematic comparison through \framework between maximum link
		utilization induced state-of-the-art algorithms relying on weight optimization
		or SR.}
	\label{fig:pie}
\end{figure}

Second, our experiments motivate questions about TE primitives and technologies.
As an illustration, we compared results obtained by IGP-WO with those returned
by our SR algorithms (keeping the best result between MILP and DEFO, for
each experiment).
Figure~\ref{fig:pie} summarizes such comparison, when all the solvers are run
for 5 minutes per experiment, on all topologies with inverse-capacity link
weight assignment.
While SR enables a substantially better solution in $25\%$ of the cases,
solutions obtained by re-weighting links and using SR are quite similar from the
congestion-avoidance perspective in most cases (within $1\%$ in $57\%$ of our
experiments), with IGP-WO being even more effective than SR algorithms for about
$7\%$ of the settings. This raises questions about the real expressive power of
segment routing, and its usefulness in (which) network setups.

\section{Conclusions}\label{sec:conclusions}

Repeatability of experiments is a key ingredient of science, and a
largely-unfulfilled need in the networking community.

This paper shows that we can overcome major obstacles to repeatability.
Prominently, we can simplify the release of code in a way that allows repeatable
evaluation and comparison with the state of the art.
We presented \framework, a software framework that
automates most of the experimental setup and evaluation process
for traffic-engineering algorithms.
To this end, \framework (i) includes a large and diversified input dataset,
(ii) implements procedures (e.g., computation of shortest paths and
traffic-to-path mapping), baselines (e.g., multi-commodity flow solutions) and
cutting-edge algorithms for traffic engineering, and (iii) includes a
pre-defined set of analyses tailored to traffic-engineering.
Our experiments confirm the practicality of \framework for both
repeating previous evaluations and performing new ones, that both deepen pros
and cons of specific algorithms and open new research questions.
We invite other researchers to join our effort, integrating our released code
and proposing similar tools for other areas.

Our work also aims at raising a serious discussion on how to work around
real-data confidentiality (e.g., for network topologies and traffic data).
Our proposal is to lower the need for proprietary data rather than keep looking
for a method to share them.
We bootstrapped a repository that we envision to collect real or realistic
network topologies for repeatable TE evaluations.
While current topologies mainly refer to specific networks (Internet
transit provider ones, at per-router level), we plan to further enrich our
repository, possibly integrating feedback from operators and vendors.

\section*{Acknowledgements}
We would like to thank Olivier Bonaventure and Renaud Hartert for initial
discussion on repeatability in networking and traffic engineering.
We are also grateful to Nikola Gvodziev for comments on a previous version of
this paper.
This work has been partially supported by the ARC grant 13/18-054 from
Communaut\'{e} fran\c{c}aise de Belgique.

\section*{Appendix: \framework LP Models}\label{appendix}

We detail here the mathematical formulation of the Linear Program (LP) models used in
\framework.

\smallskip

\myitem{General notation.}
We use $V$ and $E$ to indicate the sets of nodes and links in
the input network, respectively. We write $c(e)$ for the capacity of edge $e \in E$, and
$T_{ij}$ for the traffic volume from node $i \in V$ to $j \in V$.
Also, we use $\ecmp_{ij}$ to denote the load that shortest-path routing would generate on
each link if a demand of $1$ unit was routed from $i$ to $j$.
That is, $\ecmp_{ij}(e)$ is a fractional value in $]0, 1]$ when $e$ is on a
shortest path from $i$ to $j$, $0$ otherwise.

\smallskip

\myitem{Linear Program for theoretical optimum computation}.
\framework solves the multi-commodity flow
problem
through a linear program (LP), to provide a baseline for TE algorithms (see
\S\ref{subsec:framework-basics})
The mathematical formulation of such an LP follows: the first constraint ensures
the flow conservation in any node $i$ for the traffic destined to any node $t$,
while the second constraint defines the maximum link utilization.

\vspace{1.2em}

\noindent Variables:
\vspace{0.3em}\\
$
\begin{array}{ll}
{\bf \mathbf{load}^t(e)} & \text{fraction of the load on edge } e \text{ to destination } t\\
{\bf U}                                                                            & \text{maximum link utilization}\\
\end{array}
$
\\

\noindent Minimize ${\bf U}$ under:
\vspace{0.3em}\\
$
\begin{array}{llc}
\forall t \neq i & \sum_{e = ij \in E}{\bf \mathbf{load}^t(e)} - \sum_{e = ji \in E}{\bf \mathbf{load}^t(e)} = T_{it} \\
\forall e \in E  &  \sum_{t}{\bf \mathbf{load}^t(e)} \leq c(e) {\bf U} \\
\end{array}
$
\\

\medskip

Note that demands are aggregated by destination, which requires $O(|V||E|)$
variables, rather than $O(|V|^2|E|)$ as in per-demand models. This is enough to
compute the maximum link utilization, but does not retain per-demand information
(e.g., the fraction of link load due to each demand).

\medskip

\myitem{MILP SR model.}
\framework includes a mixed integer linear program inspired
by~\cite{belllabs-infocom15} (see \S\ref{subsec:impl-solvers}).
This MILP assumes that demands can only be re-routed via at most one detour.
This means that every path from any $i$ to any $j$ can be represented by a
variable $\srpath_{ij}$ equal to $ikj$ for some node $k$ (possibly, $k=j$ if
there is no detour).
We however reformulated the original model to have $O(|V^2E|)$ variables rather than
$O(|V^3 E|)$.
To do so, we separately group the traffic before a detour (step1) and after it
(step2).

\newpage

\noindent Variables:
\vspace{0.3em}\\
$
\begin{array}{llc}
{\bf \mathbf{path}_{ikj}} = 1 & \text{iff } \srpath_{ij} = ikj \\
{\bf s^{(1)}_{ij}} & \text{traffic to send from i to j in step 1} \\
{\bf s^{(2)}_{ij}} & \text{traffic to send from i to j in step 2}  \\
{\bf U}                                    & \text{maximum link utilization} \\
\end{array}
$
\\

\noindent Minimize ${\bf U}$ under:
\vspace{0.3em}\\
$
\begin{array}{llc}
\forall i, j              &  \sum_{k} {\bf \mathbf{path}_{ikj}} = 1  \\
\forall i, k \in V    &  {\bf s^{(1)}_{ik}} = \sum_i T_{ij} {\bf \mathbf{path}_{ikj}} \\
\forall k, j \in V    &  {\bf s^{(2)}_{kj}} = \sum_j T_{ij} {\bf \mathbf{path}_{ikj}} \\
\forall e \in E        &  \sum_{i, j} \ecmp_{ij}(e) ({\bf s^{(1)}_{ij} + s^{(2)}_{ij}}) \leq c(e){\bf U}  \\
\forall i, j, k \in V  &  {\bf \mathbf{path}_{ikj}} \in \{ 0, 1 \} \\
\end{array}
$

\end{document}